\pdfoutput=1
\documentclass[a4paper,american,floatfix,longbibliography,pdftex,superscriptaddress,twocolumn,twoside,%
aip,jcp,%
citeautoscript,
reprint
]{revtex4-1}
\usepackage{amsfonts,amsmath,amssymb}
\usepackage{dcolumn}
\usepackage[T1]{fontenc}
\usepackage[utf8]{inputenc}
\usepackage{graphicx}
\usepackage{microtype}
\usepackage[textsize=footnotesize]{todonotes}
\usepackage{hyperref,hypernat}
\usepackage[todos]{CMImacros}

\newcolumntype{d}[1]{D{.}{.}{#1}}
\newcommand{\cfeldesy}{\affiliation{Center for Free-Electron Laser Science, Deutsches
      Elektronen-Synchrotron DESY, Notkestrasse 85, 22607 Hamburg, Germany}}%
\newcommand{\ucg}{\affiliation{Department of Sciences, University College Groningen, University of
      Groningen, Groningen, Netherlands}}%
\newcommand{\ucgr}{\affiliation{Rijksuniversiteit Groningen Biomolecular Sciences and Biotechnology
      Institute, University of Groningen, Groningen, Netherlands}}%
\newcommand{\uhhcui}{\affiliation{Center for Ultrafast Imaging, Universität Hamburg, Luruper
      Chaussee 149, 22761 Hamburg, Germany}}%
\newcommand{\uhhphys}{\affiliation{Department of Physics, Universität Hamburg, Luruper Chaussee 149,
      22761 Hamburg, Germany}}%
\newcommand{\jkemail}{\email[Email:~]{jochen.kuepper@cfel.de}}%
\newcommand{\maemail}{\email[Email:~]{muhamed.amin@cfel.de}}%
\newcommand{\cmiweb}{\homepage[website:~]{https://www.controlled-molecule-imaging.org}}%

\begin{document}
\title{Variations in Proteins Dielectric Constants}
\author{Muhamed Amin}\maemail\cfeldesy\ucgr\ucg%
\author{Jochen Küpper}\jkemail\cmiweb\cfeldesy\uhhphys\uhhcui%
\date{\today}%
\begin{abstract}\noindent
   Using a new semi-empirical method for calculating molecular polarizabilities and the
   Clausius–Mossotti relation, we calculated the static dielectric constants of dry proteins for all
   structures in the protein data bank (PDB). The mean dielectric constant of more than 150,000
   proteins is $\epsilon_r=3.23$ with a standard deviation of $0.04$, which agrees well with
   previous measurement for dry proteins. The small standard deviation results from the strong
   correlation between the molecular polarizability and the volume of the proteins. We note that
   non-amino acid cofactors such as Chlorophyll may alter the dielectric environment significantly.
   Furthermore, our model shows anisotropies of the dielectric constant within the same molecule
   according to the constituent amino acids and cofactors. Finally, by changing the amino acid
   protonation states, we show that a change of pH does not have a significant effect on the
   dielectric constants of proteins.
\end{abstract}
\maketitle

\section*{TOC Graphic}
\label{sec:toc-graphic}
\begin{center}
   \includegraphics[height=50mm]{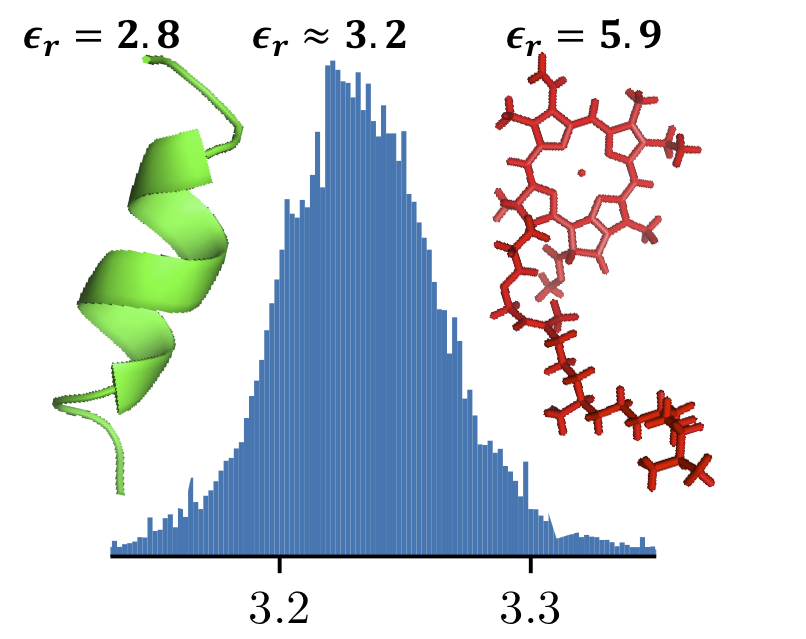} \\
   \hrulefill
\end{center}

\noindent%
The intermolecular electrostatic interactions in proteins are scaled by their dielectric constants,
which vary according to the size and composition of the proteins. The accurate determination of the
dielectric constant is essential to understand a variety of biochemical interactions such as
electron and proton transfer~\cite{Huynh:CR107:5004, Amin:JPCB:7366}, voltage
gating~\cite{Armstrong:Neuron20:371, Sigworth:QRB27:1}, ion channel
selectivity~\cite{Keramidasa:PBMB86:161}, charge separation~\cite{Gray:ARB65:537}, and
protein-protein and protein-ligand interactions~\cite{Sheinerman:10:153}. To a large extend, these
interactions are governed by the electrostatic-potential surfaces of proteins.

Direct measurements of dielectric constants $\epsilon_r$ of dry proteins span a range from 2.5 to
3.5. These values are determined by measuring the capacity of crystalline
samples~\cite{Rosen:TFS59:2178, Takashima:JPC69:4176}, which agree with chemical shift perturbation
measurements\cite{Kukic:JACS135:16968}. However, in addition to amino acids, proteins in practice
contain solvent molecules as well as organic and inorganic cofactors. These affect their dielectric
constants and in most cases the effective dielectric constant is significantly different from the
measured values for the dry proteins. The effective dielectric constants are usually determined
indirectly using the Poisson-Boltzmann equation to calculate the electrostatic interactions that
reproduce measured $pK_a$’s of some amino acids. These measurements include the effect of solvent
molecules on the dielectric constant~\cite{Kukic:JACS135:16968}. The effect of solvent on the
effective dielectric constant was studied theoretically based on Kirkwood-Fröhlich dielectric
theory~\cite{Gilson:B:2097}

In addition, computational studies based on continuum electrostatics and molecular dynamic
simulations showed that different structural motifs within the same protein may yield significantly
different values of $\epsilon_{r}$ according to the polarity of their constituents
molecules~\cite{Li:JCTC9:2126, Hazra:JMB:2282, Simonson:PNAS:1082}.

The dielectric constant $\epsilon_r$, the average polarizability $\alpha$, and the volume $V$ of a
molecule are related by the Clausius–Mossotti relation~\cite{Jansen:PR112:434}:
\begin{equation}
   \frac{4\pi\alpha}{3V} =\frac{\epsilon_{r}-1}{\epsilon_{r}+2}
   \label{eqn:clausius}
\end{equation}
However, calculations of the molecular polarizabilities of macromolecules are challenging and
computationally demanding. Previously, we proposed a model~\cite{Amin:JPCL10:2938} for calculating
the complete polarizability tensor of a protein through scaling of the tensor of a perfect conductor
of the same shape based on a molecular basis set. The scaling factor was obtained from a regression
model that correlated the polarizabilities of the molecule and a corresponding perfect-conductor of
the constituent molecules of the proteins, \ie, the amino acids.

Here, we propose a new method for the calculation of the average (scalar) polarizabilities of
proteins based on their amino acid compositions, which utilizes the fact that objects with the same
volume $V$ and dielectric constant $\epsilon_r$ have the same average polarizabilities $\alpha$
independent of shape, see also \eqref{eqn:clausius}. The static dielectric constants are then
calculated using the Clausius–Mossotti relation. This method is computationally highly efficient and
facilitated the calculations of the average polarizabilities and dielectric constants of all
proteins in the protein data bank (PDB)~\cite{Berman:NuclAcidRes28:235}.

The average polarizability of a molecule can be calculated from the sum over hybridization
configurations of the atoms in the molecule~\cite{Miller:JACS101:7213},
\begin{equation}
   \alpha = \frac{4}{N} (\sum_{A}{\tau_A})^2
   \label{eqn:alpha}
\end{equation}
with the number of electrons in the molecule $N$ and the hybrid component $\tau_A$ of each atom $A$,
obtained by approximating the zeroth order wavefunction by an antisymmetrized product of molecular
orbitals and spin functions. Average polarizabilities predicted by this method showed a very good
agreement with experimental polarizabilities for more than 400 relatively small molecules with only
$\ordsim2$~\% error.

Furthermore, since the atomic hybridizations of the atoms within the constituent amino acids do not
change in proteins, \eqref{eqn:alpha} could be rearranged to obtain the average polarizability of a
protein $\alpha_p$ by summing over effective amino-acid hybrid components:
\begin{equation}
   \label{eqn:alpha_p}
   \alpha_p =\frac{4}{N_p} (\sum_{aa}{\tau_{aa}})^2
\end{equation}
Here, $N_p$ is the number of electrons in the protein and $\tau_{aa}$ are the hybridization
components of amino acid $aa$, which are obtained as
\begin{equation}
   \label{eqn:tau}
   \tau_{aa} = \frac{\sqrt[2]{N_{aa}\alpha_{aa}}}{2}
\end{equation}
with the number of electrons $N_{aa}$ in an amino acid $aa$ and its average polarizability
$\alpha_{aa}$. The latter could be obtained from quantum-chemical calculations and, therefore, the
values of $\tau$ not only include the summation of the atomic hybrid components within the amino
acids, but also exchange correlation interactions at the level of quantum-chemistry employed.

Furthermore, for \eqref{eqn:alpha} to be applicable for very polar compounds, $\tau_{A}$ has to be
modified to include the effect of the atoms to which $A$ is bonded. However, $\tau_{aa}$ already
includes this effect since it reproduces the exact polarizabilities calculated from first
principles.

The values of $\tau$ and $\alpha_{aa}$ obtained with DFT are reported in \autoref{tab:tau} for the
6-31G+($d,p$) and 6-311G++($3df,3pd$) basis set using B3LYP functional. The 6-31G+($d,p$) basis sets
allow us to compare the predicted average polarizability against the calculated ones for the
Trp-cage mini protein, whereas the DFT calculations were not feasible for the larger basis sets. The
average polarizability of the Trp-cage protein calculated by DFT is 221~\AA$^3$; this calculation
consumed more than 2000~CPU hours. The average polarizability of Trp cage calculated with our
semi-imperial approach is 215~\AA$^3$, with an error against DFT of 2.7~\%; calculated in less than
200~\us. Thus, this approach allows the calculations the average polarizabilities and hence the
dielectric constants of all the structures stored in the PDB. However, for these calculations we
will use the amino acids polarizabilities obtained with the larger basis sets 6-311G++($3df,3pd$) to
get more accurate predictions; for Trp cage this approach yields $234~\text{\AA}^3$.

To compare with our previous method, which allows the calculations of the full polarizability
tensor, we calculated the polarizability tensor for perfect conductors of the same shape of the
proteins by solving Laplace's equation with Dirichlet boundary conditions and using Monte Carlo path
integral methods~\cite{Mansfield:PRA64:061401}. Then, all tensors are diagonalized to transform the
proteins to the polarizability frame and the average of the diagonal elements are scaled by 0.26,
which was the slope of the best-fit line that described the correlation between the amino acids and
perfect conductors of their shapes~\cite{Amin:JPCL10:2938}. The obtained polarizabilites from the
summation of the square of the atomic hybridization components highly correlate with those obtained
by scaling the polarizabilites of perfect conductors with $R^2=0.8$ and a slope of 1.6, with the
intercept set to zero. Thus, the later, previous, method produced polarizabilities that are 60~\%
higher, which we ascribe to effects of the uneven concentration of the individual amino acids in
each protein. Overall, the method presented here provides a computationally highly efficient method
for the calculation of the scalar polarizabilities. If the tensorial properties of the
polarizability are needed, the current method could be used to generate the scaling factor that is
applied to the tensor elements obtained in our previous method~\cite{Amin:JPCL10:2938}.

In order to solve the Clausius–Mossotti equation, the volumes of the proteins are calculated as the
summation of the volume of the constituent amino acids. The volume of the 20 amino acids are
calculated using the Volume Assessor software by rolling a virtual sphere with a probe radius of
1~pm on the surface of the amino acids~\cite{Voss:JMB360:893}. The calculated volumes are reported
in \autoref{tab:tau}.

\begin{figure}
   \includegraphics[width=\linewidth]{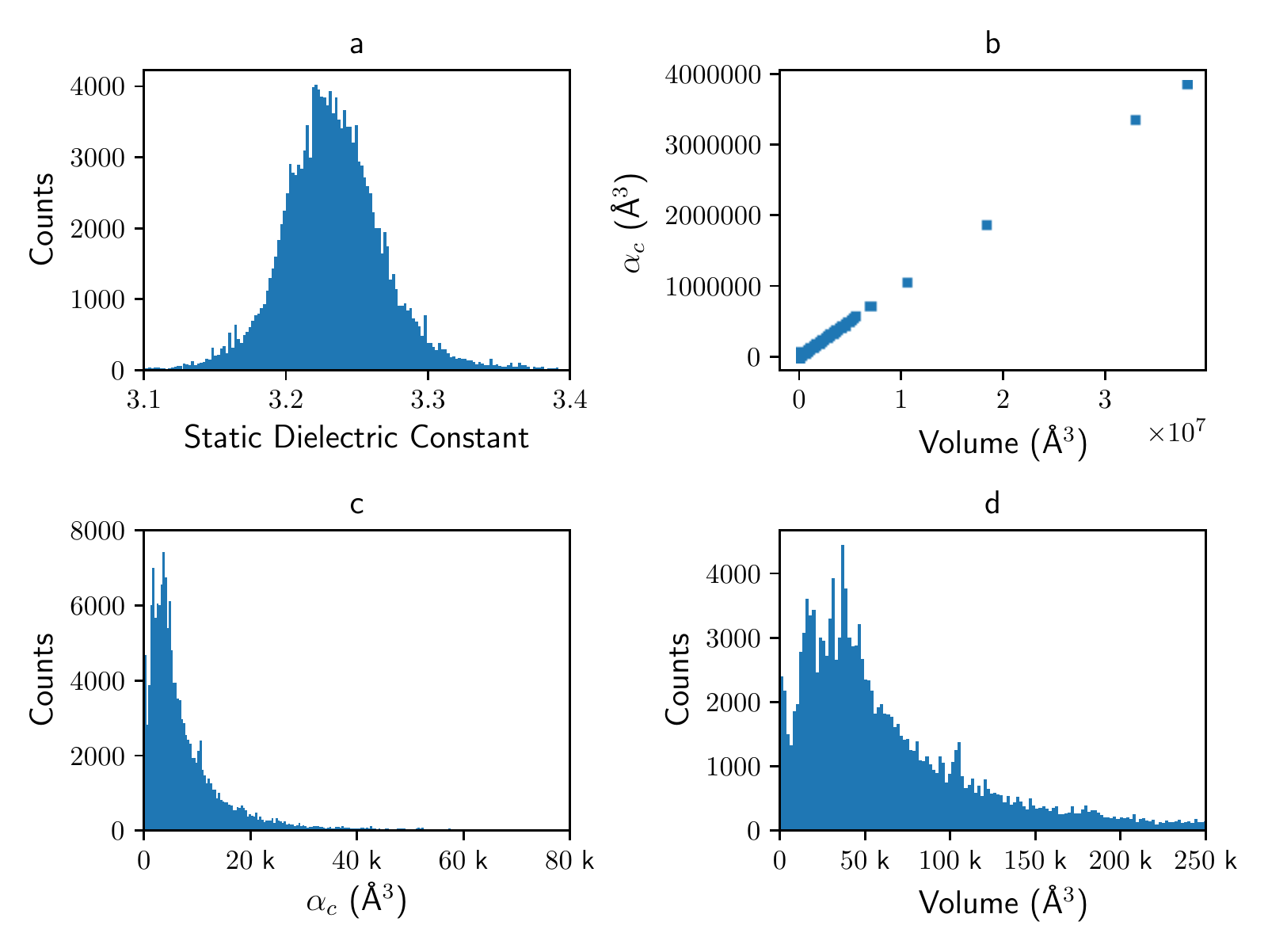}
   \caption{a) Histogram of the static dielectric constants $\epsilon_r$ of all proteins in the PDB
      database. b) Correlation between the average polarizabilities and volumes of the proteins. c)
      Histogram of the average polarizabilities of the proteins. d) Histogram of the molecular
      volume of the proteins.}
   \label{fig:correlation}
\end{figure}

\begin{table}
   \centering%
   \begin{tabular}{lccc@{\hspace{2em}}d{3.2}}
     \hline\hline
     & \multicolumn{1}{c}{$\alpha'~(\text{\AA}^3)$}
     & \multicolumn{1}{c}{$\alpha~(\text{\AA}^3)$}
     & \multicolumn{1}{c}{$V (\text{\AA}^3)$}
     & \multicolumn{1}{c}{$\frac{4\pi\alpha_l}{3V}$} \\[1pt]
     \hline
     G & 6 & 6 & 63 & 0.41 \\
     A & 7 & 8 & 81 & 0.42 \\
     S & 8 & 9 & 92 & 0.39 \\
     P & 10 & 11 & 109 & 0.41 \\
     V & 11 & 12 & 119 & 0.41 \\
     T & 10 & 11 & 109 & 0.41 \\
     C & 10 & 11 & 98 & 0.47 \\
     I & 13 & 14 & 141 & 0.41 \\
     L & 11 & 13 & 139 & 0.40 \\
     N & 10 & 11 & 112 & 0.41 \\
     D & 11 & 12 & 102 & 0.49 \\
     Q & 12 & 13 & 132 & 0.41 \\
     K & 13 & 14 & 158 & 0.36 \\
     E & 14 & 15 & 121 & 0.52 \\
     M & 14 & 15 & 139 & 0.46 \\
     H & 14 & 15 & 138 & 0.45 \\
     F & 17 & 18 & 160 & 0.48 \\
     R & 15 & 17 & 173 & 0.40 \\
     Y & 18 & 19 & 168 & 0.48 \\
     W & 22 & 23 & 193 & 0.50 \\
     \hline\hline
   \end{tabular}
   \caption{Polarizabilities, volumes, and a Clausius-Mosotti term of the amino acids. The amino acids
      are sorted b according to their molecular weight. $\alpha'$ , $\alpha$ are the average
      polarizabilities calculated using the 6-31G+($d,p$) and 6-311G++($3df,3pd$) basis sets,
      respectively. All polarizability values are reported in units of $\text{\AA}^3=\text{pm}^6$.}
   \label{tab:tau}
\end{table}

The average static dielectric constant $\epsilon_r$ for more than 150,000 protein structures stored
in the PDB database based on their amino acid decomposition is 3.23 with a standard deviation of
0.04, see \autoref[a]{fig:correlation}. According to the Clausius–Mossotti relation, the ratio
between the average polarizability and the volume, $\alpha/V$, is the factor that determines the
value of $\epsilon_r$. Thus, due to the strong correlation between the average polarizability and
the molecular volume with $R^2=1$, \autoref[b]{fig:correlation}, the standard deviation of
$\epsilon_r$ is very small. According to the regression model shown in \autoref[b]{fig:correlation},
the polarizability $\alpha$ of proteins could be calculated according to the straight line equation
$\alpha=0.1\cdot{}V+32$ with negligible residuals. Both the volume and the average polarizabilities
exhibit a skewed normal distribution, shown in \autoref[c,~d]{fig:correlation}.

The maximum dielectric constant of 3.7 is observed for N-terminal human brand 3 peptide with PDB ID
2BTA~\cite{Schneider:Biochemistry34:16574}, which has an average polarizability of
$212.7~\text{\AA}^3$ and a volume of 1879~\AA$^3$. The large polarizability of this peptide is
attributed to the ASP and GLU amino acids, which represent 50~\% of the constituent amino acids and
have high $\alpha/V$ ratios. The minimum $\epsilon_r$ of 2.8 is observed for peptide-membrane PDB ID
6HNG~\cite{Schneider:6HNG}, which is formed by only eight leucine and six lysine amino acids. The
lysine amino acid generally has a small $\alpha/V$ ratio, because it is positively charged, \ie, it
has less electrons than neutral or negatively charged amino acids which are also stronger bound.

Within the same protein the value of $\epsilon_{r}$ may change according to the composition of the
different parts. For example, in norrin, a Wnt signaling activator, PDB ID
5BPU,~\cite{Chang:Elife4:e06554} the chains A, B, D, E, and F have $\epsilon_r=3.20$, while chains H
and I have $\epsilon_r=4.26$ as they are only formed by GLU amino acids. Thus, $\epsilon_r$
distributions can be inhomogeneous within a protein, which agrees with previous studies based on MD
simulations and continuum electrostatics simulations~\cite{Li:JCTC9:2126, Hazra:JMB:2282,
   Simonson:PNAS:1082}. Furthermore, proteins have a variety of cofactor such as chlorophyll, metal
clusters, chloride ions, hems, quinones, \ldots These molecules are very different than the amino
acids and could have large impact on the dielectric environment of the proteins. For example, the
calculated average polarizability of chlorophyll is $132.3~\text{\AA}^3$, with a volume of
$900~\text{\AA}^3$, which results in $\epsilon_r=5.9$, while for iron-sulphur clusters of
photosystem~I in the oxdized state\cite{Golbeck:ARP43:293}, and its amino acids ligands
$\epsilon_r=3.2$.

To study the effect of pH on the dielectric constant, we recalculated the distribution of
$\epsilon_r$ for all proteins by replacing the average polarizabilities $\alpha_{aa}$ of GLU$^-$,
ASP$^-$, and HIS$^0$ with the average polarzbilities of the protonated form GLU$^0$, ASP$^0$, and
HIS$^+$ to simulate low pH environment. The mean of the distribution reduced to $3.15$ and the
standard deviation is unchanged. Because the mean of the $\epsilon_r$ is changed only by 0.08, it is
a reasonable assumption that proteins, which experience pH gradient across different structural
motifs have the same dielectric constants.

In conclusions, we developed an empirical method for the calculation of the average polarizabilities
of dry proteins based on their amino acids composition. The method is computationally highly
efficient and allowed us to calculate the average polarizabilities and dielectric constants of all
molecular structures in the PDB. The average dielectric constant for more than 150,000 proteins is
$\epsilon_r=3.23$, with a very small standard deviation of 0.04, due to the strong correlation
between the average polarizability and the molecular volume.

However, organic and inorganic cofactors could alter the dielectric environment of the proteins
significantly. Thus, in order to understand the chemical reactions in proteins, the correct
dielectric environment should be implemented in the biochemical/biophysical calculations.

We point out that the current approach does not take into account the molecules shape, which is
valid for the scalar average polarizability, see also \eqref{eqn:clausius}. For the computation of
tensorial properties advanced, more expensive methods have to be employed~\cite{Amin:JPCL10:2938}.

\section*{Supporting Information}
We provide a compressed text file in comma-separated-value format that contains the
polarizabilities, the volumes, and the dielectric constants for all structures in PDB (as of
01.~August 2019).

\section*{Acknowledgment}
This work has been supported by the European Research Council under the European Union's Seventh
Framework Programme (FP7/2007-2013) through the Consolidator Grant COMOTION (614507) and by the
Deutsche Forschungsgemeinschaft through the Cluster of Excellence ``Advanced Imaging of Matter''
(AIM, EXC~2056, ID~390715994).

\bibliography{string,cmi}%
\end{document}